\documentclass[12pt,preprint]{aastex}
\usepackage{bbm}
\usepackage{mathrsfs}
\usepackage{amssymb,amsmath,float}

\newcommand\de{\delta}

\newcommand\ta{\tau}

\newcommand\om{\omega}

\newcommand\De{\Delta}

\newcommand\Om{\Omega}

\newcommand\<{\langle}
\renewcommand\>{\rangle}

\newcommand\ie{\emph{i.e.}}
\newcommand\eg{\emph{e.g.}}

\newcommand\beq{\begin{equation}}
\newcommand\eeq{\end{equation}}
\newcommand\bea{\begin{eqnarray}}
\newcommand\eea{\end{eqnarray}}
\newcommand\bal{\begin{align}}
\newcommand\eal{\end{align}}

\newcommand\fr{\frac}

\newcommand\ap{\approx}

\newcommand\bk{\bold{k}}

\newcommand\br{\bold{r}}

\renewcommand\bal{\mbox{\boldmath$\alpha$}}

\newcommand\noi{\noindent}

\begin{document}

\title{A new way of detecting intergalactic baryons}

\author{Richard Lieu and Lingze Duan}

\affil{Department of Physics, University of Alabama, Huntsville, AL 35899.}

\begin{abstract}
For each photon wave packet of extragalactic light, the dispersion by line-of-sight intergalactic plasma causes an increase in the envelope width and a chirp (drift) in the carrier frequency.  It is shown that for continuous emission of many temporally overlapping wave packets with random epoch phases, such as quasars in the radio band, this in turn leads to quasi-periodic variations in the intensity of the arriving light on timescales between the coherence time (defined as the reciprocal of the bandwidth of frequency selection, taken here as of order 0.01 GHz for radio observations) and the stretched envelope, with most of the fluctuation power on the latter scale which is typically in the millisecond range for intergalactic dispersion.  Thus, by monitoring quasar light curves on such short scales, it should be possible to determine the line-of-sight plasma column along the many directions and distances to the various quasars, affording one a 3-dimensional picture of the ionized baryons in the near universe.

\end{abstract}


\noi

\section{Introduction}

After the reionization epoch, the diffuse baryons of the universe has a plasma component that for $z \lesssim 1$ redshifts is predominantly within the `warm' temperature range 10$^{5-6}$~K, and accounting for 40 -- 50 \% of the baryons in the near universe (\cite{cen99,dav01}).  This form of matter fills the entire intergalactic medium (IGM) filamentarily, with concentrations near clusters and groups of galaxies.  It still largely evades detection because the emission is in the EUV and soft X-rays, a wavelength passband that suffers from Galactic absorption as well as interstellar and time variable heliospheric foreground contamination, (see \eg~\cite{tak08} on the last; and for up to date information on the search of these `missing' baryons see \cite{bon12} and the review of \cite{dur08}).  For this reason other ingenious and `orthogonal' techniques, including and especially the recent search by the \texttt{Planck} team of `excess' Sunyaev-Zel'dovich (SZ) signals at the outskirts of clusters (\cite{maz12,arn12})
may be very important, as both papers reported positive results.  In particular, the `excess' Compton $y$-parameter at the 2 -- 5 Mpc radii of the Coma cluster
(\cite{maz12})
is consistent with the predictions based upon soft X-ray observations of Coma (\cite{lie09}), as both SZ and X-ray data yielded the same 5 Mpc limiting radius for the extent of the warm baryons as well.

Nonetheless, the warm baryons are not necessarily confined to clusters' vicinity alone, but can also structurally fill the rest of the IGM so that even the SZ technique is unable to deliver a complete search.  This is because the technique suits regions of high baryonic density and temperature, {\it viz.} baryons inside clusters can exert a large SZ `pressure' along the line-of-sight of interest, whereas outside the clusters the plasma becomes cooler and more tenuous, so that this `pressure' can easily fall below SZ detection thresholds.  The search for genuinely intergalactic baryons must therefore await new approaches.

Here we intend to discuss one of them.  Although the primary difficulty with the potentially very powerful method of utilizing the light from quasars (the only point sources bright enough to be visible across cosmological scales) as probes of the intervening universe is the lack of fast 'pulsar like' intrinsic variations in their intensity (\cite{laz08,den02}, the latter showed that rapid (hourly or shorter) variabilities in the quasar light are not intrinsic to the source), it will be argued in this Letter that the plasma dispersion effect of photon envelope broadening and carrier wave chirping are imprinted upon even steady and continuous light signals as they pass through the vast spans of the IGM.  The imprint can be uncovered by conventional techniques if the observations are done at radio frequencies.  Previous effort in this vein were mainly done with non-astrophysical applications in mind.  They include measurements\footnote{For the theory see \texttt{http://light.ece.illinois.edu/ECE460/PDF/LCI.pdf}.} of coherence length (\cite{hit99}), and calculation of the broadening of wide pulses (\cite{sal82}) and pulse distortion to 3rd order Taylor series correction (\cite{mar80}) in a dispersive medium.  To the best of the authors' knowledge there has not been any treatment of the time {\it resolved} mutual coherence function of continuous light and its associated noise characteristics in the same.

\section{The dispersion of quasar light}

We begin by revisiting the question of how, in an astrophysical context, a single wave packet of light behaves as it passes through a dispersive medium.  Since the cosmological journey of light from an unresolvable point source to a small telescope on earth is principally a one dimensional problem $\bk = (k,0,0)$ one can {\it under this scenario} ignore the dynamics of the wave packet in the $y$ and $z$ directions, \ie~the spreading of the packet is only appreciable along $x$ (scattering by plasma clumps can cause broadening, but for a mean plasma column density equivalent to 1 Gpc the effect is at the 10$^{-6}$ s level (\cite{bha04}), which as we shall see in section 3 is $\ap 10^4$ times below that of plasma dispersion).  The amplitude of a 1-D gaussian wave packet emitted at $t=t_e$ (more precisely the packet's center is $x=x_e$ at $t=t_e$) at a point $(t,x)$ `downstream', may be written as \beq \psi (t,x) = \int_{-\infty}^\infty~e^{-\fr{(k-k_0)^2}{2(\Delta k)^2}}~e^{i[k(x-x_)-\omega (t-t_e)]}~dk, \label{packet} \eeq where the spectral filter is $\propto |f(k)|^2$ with \beq f(k) = e^{-\fr{(k-k_0)^2}{2(\Delta k)^2}}, \label{spec} \eeq  and is peaked at $k_0$ of width $\De k$.  The filter $f$ usually depicts either line emission at the source or passband selection by the observer, the two are equivalent because the plasma medium does not distort the frequency spectrum of the radiation.

An approximate form of (\ref{packet}) that reveals its salient features is afforded by Taylor expanding $\omega = \om (k)$ around $k_0$ to 2nd order (3rd and higher order terms can be ignored in the case of plasma dispersion provided $\De k/k \ll$ 1),
(\ref{packet}) then becomes
\bea  \psi (t-t_e,x) &=& A \left[\fr{2\pi (\De k)^2}{1+i\om''_0 (\De k)^2 (t-t_e)}\right]^{\fr{1}{2}} \exp~[ik_0 (x-x_e)-i\omega_0 (t-t_e)] \nonumber\\ &&~\exp\left\{-\fr{(\De k)^2}{2}\fr{[x-x_e - v_g (t-t_e)]^2[1-i\om''_0 (t-t_e)(\De k)^2]}{1+\om_0^{''2} (\De k)^4 (t-t_e)^2}\right\}, \label{spread}  \eea where $v_g = (d\om/dk)_{k=k_0}$ and $\om''_0 = (d^2\om/dk^2)_{k=k_0}$.  Details of the derivation of (\ref{spread}) are given in \eg~\cite{boh51}, section 3.5.

We turn to look at the nature of continuous light from a cosmological source that is steady on short time scales.  The amplitude of which at some instance $t$ and position $\br$, with $t$ as a variable and $\br$ held fixed, may be expressed as the sum of many overlapping pulses of amplitudes $\psi_j (t)$, each depicting a single photon consisting {\it initially} of a harmonic (\ie~unchirped) carrier wave with random phase and enveloped by the coherence time $\ta_c$, {\it viz.} \ie~\beq \psi (t) = \sum_j \psi_j (t),~{\rm and}~\ta_c = \fr{1}{\De\om}. \label{psi} \eeq  In our present labeling scheme the photon pulses are chronologically ordered in terms of their envelope peaks.  The actual number $n$ of such pulses that significantly influence any given instance $t$ is finite, however, due to the finite width $\ta_c$.  This is the situation before dispersion.  In fact, $n$ is the `occupation number', or the number of arriving photons per bandwidth per coherence time that becomes an invariant for a given source and telescope.

The amplitude function of the {\it arriving} light, however, may be quite different.  It is still given by (\ref{psi}a) of course, but the constituent wave function $\psi_j$ may now be modified, {\it viz.}~\beq \psi_j (t) = \frac{A}{c}\sqrt{\fr{2\pi}{(1+i\xi)}} \De\om~\exp[-a(t-t_j)^2]~\exp[ib(t-t_j)^2-i\om_0 (t-t_j) + \phi_j], \label{psiarr} \eeq where \beq a =
\fr{(\De\om)^2}{2(1+\xi^2)},~{\rm and}~b= \xi a;~{\rm with}~\xi = \fr{\om''_0 (\De\om)^2 (t_j-t_e)}{c^2}, \label{abxi} \eeq  and $\phi_j$  a random phase\footnote{Our treatment here does {\it not} apply to the scenario of phase coherence among many temporally contiguous photons, \ie~the phenomenon of photon bunching (\eg~\cite{ric75}) which is unlikely to apply to intrinsically slow emitters like quasars.}.  Note that $\xi$ changes very slowly with time and may be treated as a constant.  In an expanding universe, it is determined by the comoving distance to the source, and the carrier frequency and average IGM plasma density (with both evaluated at the present epoch), as we shall find out.

Let us first examine the microscopic variability of the arriving intensity $I= |\psi (t)|^2$ which  may be broken down into two parts, $I = \bar I + I_1$, where \beq \bar I = \sum_{j=1}^{n'+1} |\psi_j (t)|^2,~{\rm and}~I_1 = \sum_{j\neq k}^{n'+1} \psi_j (t)\psi_k^* (t).  \label{I} \eeq  With the help of (\ref{psiarr}), one may in turn write $\bar I$ as \beq \bar I = |A|^2 \fr{2\pi}{c^2\sqrt{1+\xi^2}} (\De\om)^2 \sum_{j=1}^{n'+1} e^{-2a(t-t_j)^2} = |A|^2 \fr{\pi^{3/2}n'}{c^2\sqrt{2(1+\xi^2)}} (\De\om)^2 = |A|^2 \fr{\pi^{3/2}n}{\sqrt{2}c^2}(\De\om)^2, \label{barI} \eeq  where the rightmost expression applies to the limit of large occupation number $n \gg 1$ (hence $n' \gg 1$ necessarily) when $\bar I$ becomes a constant.

Next we work on $I_1$, which appears as \beq I_1 = I_1 (t) = |A|^2 \fr{\pi}{c^2\sqrt{1+\xi^2}} (\De\om)^2 \sum_{j \neq k}^{n'+1} e^{-a(t-t_j)^2} e^{-a(t-t_k)^2}\cos [2b(t_k-t_j)t+\varphi_{jk}], \label{I1} \eeq  where $\varphi_{jk} = -\varphi_{kj}$ changes randomly from one distinct pair of ${j,k}$ to the next.  The mean of $I_1$ obviously vanishes.  The variance is given by \beq \sigma_I^2 = |A|^2 \fr{2\pi^2}{c^4 (1+\xi^2)} (\De\om)^4 \sum_{j \neq k}^{n'+1} e^{-2a(t-t_j)^2} e^{-2a(t-t_k)^2} \<\cos^2[2b(t_k-t_j)t+\varphi_{jk}]\>. \label{sigcalc} \eeq  Again, in the large $n$ limit the double sum may be approximated by an integral, and $\<\cos^2 \> = 1/2$.  The result is \beq \sigma_I = |A|^2 \fr{\pi^{3/2}n'}{2c^2 \sqrt{2(1+\xi^2)}} (\De\om)^2 = |A|^2 \fr{\pi^{3/2}n}{2 \sqrt{2}c^2} (\De\om)^2 = \fr{\bar I}{2}. \label{sigmaI} \eeq  Thus, both $\bar I$ and $\sigma_I$ are independent of dispersion -- they are only functions of the photon occupation number $n$ and bandwidth $\De\om$ (or coherence time $\ta_c$ via (\ref{psi}b)).

The interesting question is the {\it timescale} over which the intensity $I$ undergoes the random variation about the mean $\bar I$ with the variance of (\ref{sigmaI}).  The answer comes from the $\cos [2b(t_k-t_j)t+\varphi_{jk}]$ factor of (\ref{I1}).  Under the scenario of no dispersion, \ie~ $a=(\De\om)^2/2$ and $b=0$, the factor only depends on the set of phase angles $\{\varphi_{jk}\}$, which changes to a completely different set when the time $t$ translates by ~$\sim$ one envelope width $1/\De\om$, or $\ta_c$.  Thus, in accordance with known facts (\cite{mat75}), there is a quasi-periodic variation in the intensity on the scale of one coherence time, and the Fourier Transform of $I$ would show a spike at the frequency $\sim \De\om$.

After passage through dispersion, however, the cosine factor will still vary significantly at the level of (\ref{sigmaI}) when $\{\varphi_{jk}\}$ is replaced by a completely different set, except this now happens on the timescale of one envelope width $a^{-1/2} \ap \ta_c \sqrt{2(1+\xi^2)} \gg \ta_c$; moreover there will also be milder fluctuations over {\it shorter} timescales due to the $2b(t_k - t_j)t$ part of the cosine argument.  To elaborate, in $\sum_{j,k}$ the quantity $t_k - t_j$ ranges from $\sim \ta_c$ (the assumed timing accuracy of intensity measurements) to $\sim a^{-1/2}$.  In the former end of the range $2b(t_k - t_j) \de t \to 1$ when $\de t \sim a^{-1/2}$, while in the latter end $\de t \sim \ta_c$.  Since each equal interval of $t_k - t_j$ has about the same number of photons, this means fluctuations at the level of a fraction  of (\ref{sigmaI}) exist on all scales between the coherence time $\ta_c$ and the envelope width $a^{-1/2}$, with the Fourier power  increasing towards the scale of $a^{-1/2}$ where it reaches the full level of (\ref{sigmaI}) because of the behavior of $\{\varphi_{jk}\}$.  This is the {\it observable} imprint of dispersion upon the passing light: the microscopic details of the intensity variation are very different from the scenario of a non-dispersive medium where only the characteristic timescale of $\ta_c$ exists.

We end this section by presenting a self-consistency check of the formalism of (\ref{I}) and (\ref{psiarr}).  They can also be used to demonstrate a known fact: the independence of the coherence time of continuous light, defined as the maximum time delay for observable interference fringe contrasts, on dispersion (so long as the medium does not distort the radiation energy spectrum).  The proof involves calculating the autocorrelation function (ACF) at the delay $\ta$, or $2\int dt~{\rm Re}[\psi (t)\psi^* (t +\ta)]$, from the two equations in much the same manner as above, to obtain the result \beq 2\int dt~{\rm Re}[\psi (t)\psi^* (t +\ta)] = \bar I \cos\om_0\ta ~e^{-\ta^2/(4\ta_c^2)}, \label{auto} \eeq which shows that the periodic fringe pattern $\cos\om_0\ta$ is damped exponentially away when the delay $\ta$ exceeds $\ta_c$ (the van Cittert-Zernike theorem, see \eg~section 10.4.2 of \cite{bor70}), irrespective of dispersion\footnote{This is a consequence of the Wiener-Khinchine theorem and the fact that a dispersive medium does not change the energy spectrum of the radiation, see \eg~\texttt{http://light.ece.illinois.edu/ECE460/PDF/LCI.pdf}.} as $\xi$ does not appear anywhere here.  Moreover, because in (\ref{auto}) the pattern as given by the ratio of the ACF to $\bar I$ is not a function of $n$, each photon interferes only with itself.

\section{Measurement of IGM plasma column density}


For dispersion in a cold plasma the dimensionless quantity in (\ref{abxi}c) is given by  \beq \xi = \om''_0 (\De k)^2 t = 5 \times 10^5 \left(\fr{\De\om/\om_0}{10^{-2}}\right)^2 \left(\fr{\om_0}{6 \times 10^9~{\rm rad/s}}\right)^{-1} \left(\fr{n_e}{10^{-7}~{\rm cm}^{-3}}\right) \left(\fr{\ell}{1~{\rm Gpc}}\right). \label{xi} \eeq
The central received frequency of $\nu_0 =$ 1 GHz would correspond to radio observations.   In (\ref{xi}) we ignored a small relativistic correction due to the fact that the IGM plasma at redshifts $z \lesssim$ 1 is warm.  Note also that in ({\ref{xi}) we ignored the expansion of the universe, which does not introduce significant errors for sources with $\ell \lesssim$ 1~Gpc.
If account is taken of the {\it in situ} expansion, the dimensionless parameter $\xi$ will become \beq \xi = \left(\fr{\De\om}{\om_0} \right)^2 \fr{\om_{p}^2}{\om_0} \int_{t_e}^{t} \fr{dt'}{a(t')}, \label{coxi} \eeq where $\om_{p}$ and $\om_0$ are respectively the IGM plasma frequency and the radiation frequency as evaluated at the time of observation $t$, $t_e$ is the time of emission, and $a(t)$ is the expansion parameter.  From (\ref{coxi}) it becomes apparent that, although the value of $\xi$ for sources at large distances require redshift correction, this is made simply by interpreting $\ell$ in (\ref{xi}) as the comoving distance $\int_{t_e}^{t} cdt'/a(t')$ and the other quantities in (\ref{xi}) as having assumed their values at time $t$.  This assumes of course that the evolution of $n_e$ is caused solely by the expansion.  For sources out to at least $z=1$, the warm IGM indeed maintains its constant baryonic mass fraction of $\sim$ 50 \%, \ie~additional (evolutionary) corrections are not expected to be important; see Figure 2 of \cite{cen99}.

Two features are apparent in (\ref{xi}). First, in calculating the default value of (\ref{xi}) we assumed $\De \om/\om_0 = 0.01$, which in the context of astrophysical sources is the spectral width of a narrow emission line or the telescope frequency selection.  Second, we list in Table 1 the column density of the various plasma components that extragalactic light passes through before reaching a ground based telescope; it is evident that apart from directions intercepting rich clusters of galaxies the IGM delivers the largest effect, which is why we expressed (\ref{xi}) in representative units of $n_e$ for this component.

Turning to observational strategy, we shall focus upon distant quasars, \ie~point sources, to ensure spatial coherence in the light.
From Fig. 3 of the 5,000 quasar sample of \cite{sin12} one may assume the radio spectral index of 0.6 to obtain the intrinsic luminosity density of 10$^{31}$~ergs~s$^{-1}$~Hz$^{-1}$ at 1 GHz as a representative estimate for quasars with $z \lesssim$ 0.25 ($\ell \lesssim$ 1~Gpc).  Applying a gaussian spectral filter of $\De\om/\om_0 =$ 0.01 to the arriving continuum radiation (the spectrum of which is taken as flat), one obtains $\ta_c =$ 1.6 $\times 10^{-8}$ s, $\xi \ap$ 5 $\times$ 10$^5$ from (\ref{xi}), and $a = 8 \times 10^3$~s$^{-2}$ and $b=4 \times 10^9$~s$^{-2}$ from (\ref{abxi}).  The photon `occupation number' will be $n \ap$ 20 if a quasar at 1~Gpc distance is observed by a telescope of diameter 300 m (Arecibo, it should also be mentioned that over the beam of this telescope the equivalent number of cosmic background photons is $n_{\rm CMB} \ap 3$), and indicates that one is in the classical (phase) noise limit for the radiation intensity, see the previous section. An exposure time of 1 hour would yield $N \ap 4 \times 10^{12}$ as the number of collected photons.

Thus the numbers all point to the validity of the classical description of radiation given in the previous section, which also presented a relatively straightforward means of using the quasar light to measure line-of-sight dispersion, {\it viz.}~the microscopic random fluctuations should have their highest amplitude on the timescale\footnote{In addition to having a sufficient number of photons per such an interval, radio observations at this timing resolution is also not expected to be a problem, since the highest resolution ever reached is 4 nanosecond (\cite{han03}).} $a^{-1/2} \ap 10$ millisecond and {\it not} the (much shorter) coherence time $\ta_c$, although the Fourier transform of the light curve should reveal fluctuation power on all scales between $a^{-1/2}$ and $\ta_c$, monotonically falling from the former to the latter.  By observationally determining $a^{-1/2}$ in this way, the line-of-sight IGM plasma column may then be inferred from (\ref{abxi}a), (\ref{xi}), and knowledge of the bandwidth $\De\om$ of frequency selection.  The crucial datum here is the upper cutoff timescale of $a^{-1/2}$, \ie~one does not expect any significant variations in the intensity above this scale, until the scale $\sim 1$ day when genuine changes in the emission properties of the quasar source have been noted by \eg~\cite{egg20}.  To the best of the authors' awareness no attempt have been made to monitor quasars on the millisecond time frame.  Of course, if source variability occurs down at this range the proposed technique will fail.  It is extremely difficult to envisage such phenomena, however, because of the size of the quasar emission region.

\section{Acknowledgment}

The authors thank T.W.B. Kibble and J.F. McKenzie for helpful discussions.



\begin{table}
    \begin{tabular}{|l|l|l|l|}
        \hline
        Type of plasma            & $n_e$ (in cm$^{-3}$) & Column length $\ell$  & Column density $n_e \ell$ (cm$^{-2}$) \\ \hline
        IGM                       & 10$^{-7}h_{0.7}$                             & 3~Gpc                 & 9 $\times$ 10$^{20}$                  \\
        Rich clusters of galaxies & 10$^{-3}$                                    & 2~Mpc                 & 6 $\times$ 10$^{21}$                  \\
        Interstellar medium (ISM) & 0.03                                         & 1~kpc                 & 9 $\times$ 10$^{19}$                  \\
        Interplanetary medium (IPM) & 10   & 100 AU  & 1.5 $\times$ 10$^{16}$  \\
        Earth's ionosphere        & 10$^5$                                       & 300~km                & 3 $\times$ 10$^{12}$                  \\
        \hline
    \end{tabular}
    \caption{Properties of plasmas on various length scales, with the following noteworthy points,  On the ISM (interstellar medium) the reader should beware of the anisotropy in $n_e$ presented by the Galactic disk, \ie~the column density as tabulated remains representative of any direction including high Galactic latitudes (see \eg~\cite{how06}), assuming that one avoids the disk when performing cosmological observations.  On the IGM, the density $n_e$ is obtained by assuming that the baryonic IGM consists principally of the 10$^{5-7}$~K plasma (the WHIM, or warm-hot intergalactic medium) of section 1, with normalized cosmic density $\Om_{\rm WHIM} \ap$ 0.02 between $z=0$ and at least $z=1$ (\cite{cen99}, Fig. 2b), where $h_{0.7}$ denotes the present Hubble constant $H_0$ in units of 70~km~s$^{-1}$~Mpc$^{-1}$ and $\ell$ is the comoving distance $\int dt/a(t)$.}
\end{table}

\end{document}